\documentclass[aps,prd,preprint]{revtex4-2}
\pdfoutput=1

\usepackage[T1]{fontenc}
\usepackage[utf8]{inputenc}
\usepackage{lmodern}

\usepackage{graphicx,amssymb,amsmath,amsfonts}
\usepackage{amsthm}
\usepackage[greek, english]{babel}
\usepackage{hyperref}
\usepackage{cleveref}

\usepackage{braket}
\usepackage{xcolor}

\usepackage{mathrsfs}
\usepackage{slashed}

\newcommand{\mn}{{\mu\nu}}
\newcommand{\rs}{{\rho\sigma}}
\newcommand{\ab}{{\alpha\beta}}
\newcommand{\mnrs}{{\mu\nu\rho\sigma}}

\newcommand{\V}{\mathbf}

\newcommand{\p}{{\partial}}
\newcommand{\w}{{\omega}}

\newcommand{\vep}{\varepsilon}
\newcommand{\ep}{\epsilon}

\newcommand{\zb}{{\bar{z}}}
\newcommand{\wb}{{\bar{w}}}
\newcommand{\gzz}{{\gamma_{z\bar{z}}}}
\newcommand{\guzz}{{\gamma^{z\bar{z}}}}

\newcommand{\mI}{\mathcal{I}}

\newcommand{\mL}{\mathcal{L}}

\newcommand{\scri}{\mI}

\newcommand{\im}{\mathrm{Im}}
\newcommand{\re}{\mathrm{Re}}

\newcommand{\td}[1]{\widetilde{d^3 #1}\,}

\newcommand{\hc}{\text{h.c.}}
\newcommand{\cc}{\text{c.c.}}

\newcommand{\tin}{{\text{in}}}
\newcommand{\tout}{{\text{out}}}

\let\tilde\relax
\newcommand{\tilde}{\widetilde}

\newcommand{\rrangle}{\rangle\mkern-4mu\rangle}

\newcommand{\kket}[1]{{\left\Vert #1\right\rrangle}}

\makeatother

\begin{document}

\title{Magnetic Soft Charges, Dual Supertranslations and 't Hooft Line Dressings}
\author{Sangmin Choi}
\email{sangminc@umich.edu}
\author{Ratindranath Akhoury}
\email{akhoury@umich.edu}

\affiliation{Leinweber Center for Theoretical Physics, \\
Randall Laboratory of Physics, Department of Physics,\\
University of Michigan, Ann Arbor, MI 48109, USA}

\begin{abstract}
We construct the Faddeev-Kulish asymptotic states in a quantum field theory of electric and magnetic charges. We find that there are two kind of dressings: apart from the well known (electric) Wilson line dressing, there is a magnetic counterpart which can be written as a 't Hooft line operator. The 't Hooft line dressings are charged under the magnetic large gauge transformation (LGT), but are neutral under electric LGT. This is in contrast to the Faddeev-Kulish dressings of electrons, which can be written as a Wilson line operator and are charged under electric LGT but neutral under magnetic LGT. With these dressings and the corresponding construction of the coherent states, the infrared finiteness of the theory of electric and magnetic charges is guaranteed. Even in the absence of magnetic monopoles, the electric and magnetic soft modes exhibit the electromagnetic duality of vacuum Maxwell theory. Using only the asymptotic form of three-point interactions in a field theory of electric and magnetic charges, we show that the leading magnetic dressings, like the leading electric ones, are exact in the field theory of electric and magnetic charges, in accordance with a conjecture of Strominger. We then extend the construction to perturbative quantum gravity in asymptotically flat spacetime, and construct gravitational 't Hooft line dressings that are charged under dual supertranslations. The duality in the quantum theory between the electric and magnetic soft charges and their dressings is thus made manifest.
\end{abstract}

\maketitle

\section{Introduction}\label{sec:introduction}

The electromagnetic duality of the vacuum Maxwell theory is broken in  quantum electrodynamics with only electric sources. However, a recent work
	\cite{Freidel:2018fsk} has shown that 
this duality is regained, even in the absence of magnetic monopoles, for the asymptotic states which include the soft electric and magnetic modes.
In a theory with 
electrically and magnetically charged particles, Strominger \cite{Strominger:2015bla} has obtained the magnetic corrections to the usual soft photon theorems and explicitly constructed the charges which 
generate the magnetic large gauge transformations. A number of interesting questions, however, need to be further explored. Is the theory of electric and magnetic charges infrared finite? Are Strominger's generalized soft photon theorems exact?
The extension of supertranslations in gravity to include dual supertranslations has been studied in
	\cite{Godazgar:2018dvh,Godazgar:2018qpq,Kol:2019nkc,Godazgar:2019dkh,Godazgar:2019ikr,Huang:2019cja} and similar questions 
are relevant here as well. The aim of this paper is to address these issues. Our approach will be  to construct the dressings which are charged under magnetic LGT  in electrodynamics and under dual supertranslations in perturbative quantum gravity. The construction of dressed states to cure infrared divergences has a long history.
The dressed states of Faddeev and Kulish \cite{Kulish:1970ut} were derived as the asymptotic states of QED
		that correctly take into account the non-vanishing interactions between photons and electrons at the infinity
	and therefore cures infrared divergences of the matrix elements.
Following Mandelstam's formulation of QED \cite{Mandelstam:1962mi} without gauge potential,
	Jakob and Stefanis \cite{Jakob:1990zi} showed that the Faddeev-Kulish dressings can be written as a Wilson line operator.
The Faddeev-Kulish states of perturbative quantum gravity has been constructed in \cite{Ware:2013zja},
	and the dressing also can be written as a gravitational Wilson line operator.
Using the Wilson line interpretation, one can extend the dressing construction to curved spacetimes \cite{Choi:2018oel,Choi:2019rlz}
	where it is difficult to apply the original approach using an asymptotic interaction Hamiltonian. 

In 2013, Strominger \cite{Strominger:2013jfa,He:2014laa} pointed out that BMS supertranslation constrains
	the S-matrix of perturbative quantum gravity in a non-trivial way:
	demanding the conservation of BMS charge in scattering processes reproduces Weinberg's soft graviton theorems.
This analysis was then also applied to QED \cite{He:2014cra,Kapec:2015ena}, where it was shown the gauge transformations that do not vanish at infinity,
	referred to as large gauge transformations (LGT), have a charge conservation law that is encoded in the soft photon theorems.
It was later shown that the asymptotic states of Faddeev and Kulish shift the soft LGT charge of the vacuum in QED \cite{Gabai:2016kuf}
	(soft BMS charge in gravity \cite{Choi:2017bna}), in such a way that facilitates LGT (BMS) charge conservation.
In this perspective, the Faddeev-Kulish dressings can be understood as operators that carry a definite charge of the
	asymptotic symmetry \cite{Kapec:2017tkm,Choi:2017ylo,Choi:2018oel,Choi:2019rlz}.

This line of investigation has also been applied, as mentioned above, to theories with electric and magnetic charged particles (and dyons)
	\cite{Strominger:2015bla}.
In such theories there are two conserved charges in a scattering processes, namely the electric and magnetic LGT charges.
Even in the absence of magnetic monopoles, the magnetic soft charges are still conserved in scattering processes \cite{Campiglia:2016efb}.
In gravity, this is closely related to an asymptotic symmetry referred to as the dual supertranslations,
	which has recently gained attention as the ``magnetic'' counterpart of BMS supertranslations
	\cite{Godazgar:2018dvh,Godazgar:2018qpq,Godazgar:2019dkh,Godazgar:2019ikr,Kol:2019nkc,Huang:2019cja}.

In this paper, to resolve some of the questions raised earlier, we study the magnetic counterparts of the Faddeev-Kulish dressings.
First, working within the framework of a quantum field theory of magnetic and electric charges formulated by Blagojevi\'c and collaborators
	\cite{Blagojevic:1978zv,Blagojevic:1979bm,Blagojevic:1985sh}, we construct the asymptotic states of the magnetically charged particles
	by diagonalizing the asymptotic three point interaction potential of these with the photon.
We emphasize that the field theory formulation of \cite{Blagojevic:1978zv,Blagojevic:1979bm,Blagojevic:1985sh}
	is used only in the spirit of an effective field theory to determine the structure of the asymptotic three-point interaction.
This construction at large times is non-perturbative, since the states can also be derived by other non-perturbative methods,
	such as writing Wilson line dressings or building
	eigenstates of the soft charge associated with the asymptotic symmetry.
	These methods give identical results (see \cite{Jakob:1990zi,Kapec:2017tkm,Choi:2017ylo}).
	Later in the paper, we use only the second method for gravity.
Having obtained the asymptotic states,
	we then show that the soft photon dressing associated to these states can be written as a 't Hooft line operator
	along the asymptotic trajectory of the magnetically charged particle.
By direct computation, we demonstrate that the 't Hooft line dressings are charged under the magnetic LGT while neutral under electric LGT.
This is in contrast to the dressings for electrons \cite{Kulish:1970ut}, which can be written as a Wilson line \cite{Jakob:1990zi}
	and are charged under electric LGT while neutral under magnetic LGT. The construction of the 't Hooft line dressing parallels the treatment of electrically charged particles and 
	this acts on the Fock states to create coherent states. The infrared finiteness of the quantum field theory of electric and magnetic charges is then manifest. The construction also makes clear
	that the leading magnetic dressings, just like their electric counterparts, are exact as was conjectured by Strominger in \cite{Strominger:2015bla}.
The 't Hooft line interpretation of the dressing allows us to extend the construction to perturbative quantum gravity.
We construct gravitational 't Hooft line dressings that are charged under dual supertranslations but carry zero BMS supertranlsation charge.
Again, this is to be contrasted with gravitational Wilson line dressings, which are charged only under supertranslations.
In gravity, we have no magnetic counterpart of the graviton coupling to the energy-momentum tensor.
Thus, there are no particles that carry dual supertranslation charge, and hence no issue regarding infrared divergences.
Our aim with the construction of gravitational 't Hooft line dressings is not to resolve infrared divergences, but to study the
	gravitational analog of electromagnetic duality for the zero modes as proposed in \cite{Freidel:2018fsk}.
For this purpose, we study the algebra of dual supertranslation charges and the 't Hooft line dressings for smooth parameter functions
	on the sphere.
We expect that the extension to non-smooth parameter functions in gravity will be similar to that found in \cite{Freidel:2018fsk} for
	electromagnetism.

The paper is organized as follows.
Section \ref{sec:review} reviews the quantum field theory of magnetic and electric charges of Blagojevi\'c and collaborators.
In section \ref{sec:dressing} we construct the dressing, and demonstrate that it is a 't Hooft line operator along a given asymptotic trajectory,
	and that it only carries magnetic LGT charge.
In section \ref{sec:gravity} we extend the previous construction to gravity, to construct gravitational 't Hooft line operators.
We show that such operators are only charged under dual supertranslations.
We conclude with a discussion of these results in \ref{sec:discussion}.

\section{Quantum field theory of electric and magnetic charges}\label{sec:review}

In this section, we briefly review the one-potential Lagrangian formulation of the quantum field theory of electric and magnetic charges
	by Blagojevi\'c and collaborators \cite{Blagojevic:1978zv,Blagojevic:1979bm}.
This theory has been shown in \cite{Blagojevic:1978zv,Blagojevic:1979bm} to be equivalent to two of the more well-known
	quantum field theories of electric and magnetic charges,
	namely the Hamiltonian formulation of Schwinger \cite{Schwinger:1966nj} and the Lagrangian formulation of Zwanziger \cite{Zwanziger:1970hk}.
Our interest in this theory is not to calculate amplitudes but to understand the structure of the three-point interactions at large times.
This is all that is needed to construct the dressings.
Once the dressings are constructed, the soft theorems and the coherent states follow straightforwardly.

The formulation is based on the Lagrangian
\begin{align}
	\mL &= -\frac{1}{4}F_\mn F^\mn + \bar\psi\left[\gamma^\mu(i\p_\mu - eA_\mu)-m_\psi\right]\psi
		+ \bar\chi(\gamma^\mu i\p_\mu - m_\chi)\chi,
	\label{Lagrangian1}
\end{align}
where $\psi$ and $\bar\psi$ ($\chi$ and $\bar\chi$) are the fermionic fields describing an electrically (magnetically) charged spin-1/2 particle,
	and the field strength tensor, following Dirac \cite{Dirac:1948um}, is defined as
\begin{align}
	F_\mn &= \p_\mu A_\nu - \p_\nu A_\mu + \ep_\mnrs G^\rs,
\end{align}
with $A_\mu$ the photon field and $\ep_\mnrs$ the totally antisymmetric tensor.
Here $G^\rs$ is a functional of $\chi$ and $\bar\chi$, defined as
\begin{align}
	G_\mn(x) = \int d^4y\, h_\mu(x-y) j^g_\nu(y),
\end{align}
where $j^g_\nu = g\bar\chi\gamma_\nu\chi$ is the magnetic current, $g$ is the magnetic charge,
	and $h_\mu(x)$ is any c-number function satisfying $\p_\mu h^\mu(x) = -\delta(x)$.
A convenient form of $h_\mu$ can be given in terms of an arbitrary real vector $n^\mu$,
\begin{align}
	h_\mu(x) = -n_\mu (n\cdot \p)^{-1}(x),
\end{align}
which comes in handy for concrete calculations.
A coordinate space representation of $(n\cdot \p)^{-1}(x)$ in terms of step functions can be found in \cite{Blagojevic:1978zv};
	we will not write it here since we'll mostly be working in momentum space.

This Lagrangian yields the following Maxwell's equations,
\begin{align}
	\p^\mu F_\mn &= j_\nu^e,\qquad\\
	\p^\mu (*F)_\mn &= j_\nu^g,\qquad
\end{align}
where $j_\nu^e=e\bar\psi\gamma_\nu\psi$ ($j^g_\nu = g\bar\chi\gamma_\nu\chi$) is the conserved electric (magnetic) current, and
\begin{align}
	(*F)_\mn = \frac{1}{2}\ep_\mnrs F^\rs
	\label{Fdual}
\end{align}
is the tensor dual of $F_\mn$.

One can read off the momentum space Feynman rules from the Lagrangian.
Of interest to us is the 3-point vertex; all momenta are flowing into the vertex:
	\begin{align}
		\vcenter{\hbox{\includegraphics[width=.25\textwidth]{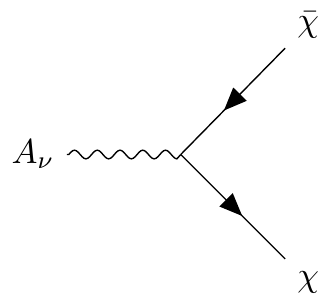}}}
		\quad=-ig\ep_\mnrs k^\mu \frac{n^\rho \gamma^\sigma}{n\cdot k}
		\equiv -igA_{\nu\sigma}\gamma^\sigma,
	\end{align}
Here we defined $A_\mn = \ep_\mnrs \frac{n^\rho k^\sigma}{n\cdot k}$.
The divergence appearing when $n\cdot k=0$ is spurious, see \cite{Blagojevic:1981he}.
Other standard Feynman rules such as the standard propagators and the electron-photon vertex have been omitted and can be found in \cite{Blagojevic:1981he}.
It is worth noting that while the Lagrangian is non-local in coordinate space, the Feynman rules are local in momentum space.

\section{'t Hooft line dressing}\label{sec:dressing}

In this section, we construct the infrared-finite dressed state of an asymptotic magnetically charged particle, along the same lines as Faddeev and Kulish
	\cite{Kulish:1970ut,Ware:2013zja}, and show that the dressing can be written as a 't Hooft line dressing.

\subsection{Faddeev-Kulish construction}\label{sec:FKconstruction}

In order to construct the dressed state, we write out the Lagrangian \eqref{Lagrangian1} as
\begin{align}
	\mL &=
		-\frac{1}{4}(\p_\mu A_\nu - \p_\nu A_\mu)
		-j_e^\mu A_\mu
		-\frac{1}{2}n^2j_g^\mu\left(g_\mn - \frac{n_\mu n_\nu}{n^2}\right)(n\cdot \p)^{-2}j_g^\nu
		\nonumber\\&\quad
		-\ep^\mnrs(n\cdot \p)^{-1}\p_\mu A_\nu n_\rho j_\sigma^g
		+ \bar\psi(i\slashed\p-m_\psi)\psi
		+ \bar\chi(i\slashed\p-m_\chi)\chi,
	\label{Lagrangian2}
\end{align}
where we recall that $n^\mu$ is an arbitrary 4-vector, $j_e^\mu = e\bar\psi\gamma^\mu\psi$ and $j_g^\mu = g\bar\chi\gamma^\mu\chi$.
We observe that the normal-ordered interaction potential relevant to the $\chi$-photon scattering is
\begin{align}
	V_{A\bar\chi\chi}(t) = g\ep^{\mnrs}\int d^3x\,
		(n\cdot \p)^{-1}:\p_\mu A_\nu n_\rho \bar\chi\gamma_\sigma\chi:.
	\label{potential1}
\end{align}
The photon and $\chi$ fields have the standard mode expansions
\begin{align}
	A_\mu(x) &=
		\int\td{k}
		\left(
			\ep^{\alpha *}_\mu(\V k)a_\alpha(\V k)e^{ik\cdot x}
			+\ep^\alpha_\mu(\V k)a_\alpha^\dagger(\V k)e^{-ik\cdot x}
		\right),
	\label{A}
	\\
	\chi(x) &=
		\int \td{p}
		\left(
			b_s(\V p) u^s(\V p)e^{ip\cdot x}
			+ d_s^\dagger(\V p) v^s(\V p)e^{-ip\cdot x}
		\right),
	\\
	\bar\chi(x) &=
		\int \td{p}
		\left(
			d_s(\V p)\bar v^s(\V p)e^{ip\cdot x}
			+ b_s^\dagger(\V p)\bar u^s(\V p)e^{-ip\cdot x}
		\right),
\end{align}
where we employ the usual notation for the Lorentz invariant measures,
\begin{align}
	\td{k} \equiv \frac{d^3k}{(2\pi)^3}\frac{1}{2\w},
	\qquad
	\td{p} \equiv \frac{d^3p}{(2\pi)^3}\frac{1}{2E_p},
\end{align}
	$\ep_\mu^\alpha(\V k)$ is the photon polarization vector,
	$u^s(\V p)$ ($v^s(\V p)$) is the fermion (anti-fermion) spinor amplitudes,
	$\w=\left|\V k\right|$, $E_p=\sqrt{\V p^2+m_\chi^2}$,
	and the creation and annihilation operators satisfy the appropriate commutation/anti-commutation relations
\begin{align}
	\left[a_\alpha(\V k), a_\beta^\dagger(\V k')\right] &= \delta_\ab(2\pi)^3(2\w)\delta^{(3)}(\V k-\V k'),
	\label{comm1}
	\\
	\left\{b_r(\V p), b_s^\dagger(\V q)\right\} &= \delta_{rs}(2\pi)^3(2E_p)\delta^{(3)}(\V p-\V q),
	\label{comm2}
	\\
	\left\{d_r(\V p), d_s^\dagger(\V q)\right\} &= \delta_{rs}(2\pi)^3(2E_p)\delta^{(3)}(\V p-\V q).
	\label{comm3}
\end{align}
Plugging in the expansions to \eqref{potential1} and taking the large-time limit $|t|\to \infty$,
	we arrive at the asymptotic potential
\begin{align}
	V^\text{as}_{A\bar\chi\chi}(t) &=
		g\ep^{\mu\nu\rho\sigma}\int \td{k}\td{p}\frac{p_\nu}{E_p}
		\frac{n_\rho k_\sigma}{n\cdot k}
		\rho(\V p)
		\left(
			\ep^\alpha_\mu(\V k)a_\alpha^\dagger(\V k)
				e^{-i\frac{p\cdot k}{E_p}t}
			+\ep^{\alpha *}_\mu(\V k)a_\alpha(\V k)
				e^{i\frac{p\cdot k}{E_p}t}
		\right),
	\label{Vas}
\end{align}
where $\rho(\V p)=\sum_s \left(b_s^\dagger(\V p)b_s(\V p) - d_s^\dagger(\V p)d_s(\V p)\right)$ is the number density operator of the magnetically
	charged particle.
We note that this asymptotic form of the 3-point interaction is all we will need of the original theory, \eqref{Lagrangian1},
	to construct dressings, coherent states and soft theorems.
One could argue the form of this interaction from symmetry arguments in the spirit of an effective field theory.
Following the same line of arguments as in the original Faddeev-Kulish construction \cite{Kulish:1970ut},
	one obtains from this potential the asymptotic state of the magnetically charged particles,
\begin{align}
	\kket{\V p_1,\ldots,\V p_n} = e^{\tilde R}\ket{\V p_1,\ldots,\V p_n},
\end{align}
with the dressing
\begin{align}
	e^{\tilde R} &=
		\lim_{t\to \infty}
		\exp\left(i\int^t V^\text{as}_{A\bar\chi\chi}(\tau)d\tau\right)
	\\ &=
		\exp\left\{
		-g\ep^{\mu\nu\rho\sigma}\int \td{k}\td{p}\frac{p_\nu}{p\cdot k}
		\frac{n_\rho k_\sigma}{n\cdot k}
		\rho(\V p)
		\phi(\w)
		\left(
			\ep^\alpha_\mu(\V k)a_\alpha^\dagger(\V k)
			-
			\ep^{\alpha *}_\mu(\V k)a_\alpha(\V k)
		\right)
		\right\}
	\\ &=
		\exp\left\{
		-g\int \td{k}\td{p}\frac{A^\mn p_\nu}{p\cdot k}
		\rho(\V p)
		\phi(\w)
		\left(
			\ep^\alpha_\mu(\V k)a_\alpha^\dagger(\V k)
			-
			\ep^{\alpha *}_\mu(\V k)a_\alpha(\V k)
		\right)
		\right\},
	\label{dressing}
\end{align}
where $A_\mn = \ep_\mnrs \frac{n^\rho k^\sigma}{n\cdot k}$.
Note that the factors $\exp(\pm i(p\cdot k) t/E_p)$ in $V^\text{as}_{A\bar\chi\chi}(t)$ at large $t$ suppress $\w>0$ contributions to the integral
	by the Riemann-Lebesgue lemma, so we have replaced such factors with an infrared function $\phi(\w)$ that only has support in a small
	neighborhood of $\w=0$ and satisfies $\phi(0)=1$ to reflect this;
	see \cite{Kulish:1970ut, Ware:2013zja} for instance.
Due to the number density operator $\rho(\V p)$, the dressing decomposes into a product of single particle dressings $e^{\tilde R} = \prod_i e^{\tilde R(p_i)}$,
	where
\begin{align}
	e^{\tilde R(p)} &=
		\exp\left\{
		-g\int \td{k}\frac{A^\mn p_\nu}{p\cdot k}
		\phi(\w)
		\left(
			\ep^\alpha_\mu(\V k)a_\alpha^\dagger(\V k)
			-
			\ep^{\alpha *}_\mu(\V k)a_\alpha(\V k)
		\right)
		\right\}.
	\label{singleFK}
\end{align}

For physical states $\Psi$, the Gupta-Bleuler condition demands
	$k^\mu a_\mu(\V k) \ket{\Psi} = 0$,
where $a_\mu(\V k) = \ep^{\alpha*}_\mu(\V k)a_\alpha(\V k)$.
The consistency of the dressing \eqref{singleFK} with this condition boils down to the commutator
\begin{align}
	0 &=
		\left[
			\frac{A^\mn p_\nu}{p\cdot k}\left(a_\mu^\dagger(\V k)- a_\mu(\V k)\right)
			,
			k'^\rho a_\rho(\V k')
		\right]
	\\&=
		\frac{k^\mu A_\mn p^\nu}{p\cdot k} (2\pi)^3(2\w)\delta^{(3)}(\V k-\V k'),
\end{align}
which is is automatically satisfied since $k^\mu A_\mn=0$ for any choice of $n$ by antisymmetry of $\ep_\mnrs$.
Therefore, the dressing commutes with the Gupta-Bleuler condition.
In the original construction of dressings in QED \cite{Kulish:1970ut}, Faddeev and Kulish introduced a vector $c^\mu$
	into the dressing to make it compatible with gauge fixing; here we do not need such a treatment.
It was recently shown by Hirai and Sugishita \cite{Hirai:2019gio} that a careful BRST analysis removes
	the need for $c^\mu$ even in the Faddeev-Kulish construction.
An analogous BRST analysis of the theory including magnetically charged particles is left for future investigation.

As the construction of the dressing \eqref{dressing} was fairly parallel to that of \cite{Kulish:1970ut}, it is natural to expect
	that it resolves the infrared divergence of the theory \eqref{Lagrangian1}.
To see this, let us consider a generalization of \eqref{Lagrangian1} to a theory containing dyons, which is fairly straightforward
	(see for instance \cite{Blagojevic:1985sh}.
In the generalized theory, dyons will be dressed with the magnetic dressing \eqref{dressing} as well as the original Faddeev-Kulish dressing of QED
	\cite{Kulish:1970ut}, where the latter takes the form
\begin{align}
	e^{R} = \exp\left\{
			-e\int \td{p}\td{k}\rho(\V p)\phi(\w)
			\frac{p^\mu}{p\cdot k}
			\left(
				a_{\mu}^\dagger(\V k) - a_\mu(\V k)
			\right)
	\right\}.
\end{align}
Since $\left [R,\tilde R\right ]=0$, the dressing in a dyonic generalization of \eqref{Lagrangian1} will take the form
\begin{align}
	\exp\left\{
		-\int \td{p}\td{k}\rho(\V p)\phi(\w)
		\left(
			e \eta_\mn + g A_\mn
		\right)
		\frac{p^\nu}{p\cdot k}
		\left(
			a^{\mu\dagger}(\V k) - a^\mu(\V k)
		\right)
	\right\},
	\label{dyon}
\end{align}
where $\rho(\V p)$ is now a number density operator of the dyon field.
When acted on a single dyon state, this dressing reproduces the coherent state of Antunovi\'c and Senjanovi\'c \cite{Antunovic:1984td}
	that was shown to resolve the infrared divergences, as expected.
Thus, the dressings we have constructed will ensure the infrared finiteness of any theory whose asymptotic three-point
	interaction is given by \eqref{Vas}.

In the next subsection, we will see that this dressing can be written as a 't Hooft line dressing along the timelike trajectory of an asymptotic
	particle with momentum $p$.

\subsection{Dressing as 't Hooft line}\label{sec:dressing_as_tHooft_line}

We will now re-derive the magnetic Faddeev-Kulish dressing \eqref{singleFK} by considering a 't Hooft line operator.
This is to be contrasted with the usual (electric) Faddeev-Kulish dressing \cite{Kulish:1970ut,Choi:2018oel}
	having a Wilson line representation.

Consider a 't Hooft operator $\exp\left(ig \int_S *F\right)$ associated with a simple connected 2-dimensional surface $S$ with boundary loop $C$.
If we can write $*F = d\tilde A$ for a vector field $\tilde A$ in the absence of electrons, by Stoke's theorem, the 't Hooft operator
	becomes $\exp\left(ig\oint_C \tilde A\right)$.
Then, we can consider the field $\chi$ dressed with a 't Hooft line,
\begin{align}
	\exp\left(ig\int^x_\infty d\xi^\mu \tilde A_\mu(\xi)\right)\chi(x).
	\label{full}
\end{align}
For an asymptotic particle, this is possible and we will see that the dressing of the asymptotic particle agrees with \eqref{singleFK}.
We will concern ourselves with large time dynamics, so in what follows below we will consider free equations of motion with $j_g=j_e=0$.

To this end, we write the Minkowski metric in the retarded time coordinates,
\begin{align}
	ds^2 = -du^2 + 2dudr + 2r^2\gzz dz d\zb,
\end{align}
where $u=t-r$ is the retarded time suitable for describing the future null infinity $\scri^+$, $z=e^{i\phi}\cot(\theta/2)$ is the complex angular coordinate,
	and $\gzz=2/(1+z\zb)^2$ is the unit 2-sphere metric.
The radiative mode of photon on $\scri^+$ can be expanded as \eqref{A},
	where we sum over the two physical polarizations $\alpha=\pm$.
In terms of the complex angular coordinates $(z_k,\zb_k)$, the $(t,x,y,z)$ components of the photon momentum $k$ becomes
\begin{align}
	k^\mu = \frac{\w}{1+z_k\zb_k}\left(1+z_k\zb_k, \zb_k+z_k, i(\zb_k-z_k), 1-z_k\zb_k\right),
	\label{photon_momentum}
\end{align}
where $\w=|\V k|$.
Then the two transverse polarization vectors can be defined as
\begin{align}
	\ep^{+\mu}(\V k) = \frac{1}{\sqrt 2}(\zb_k,1,-i,-\zb_k),
	\qquad
	\ep^{-\mu}(\V k) = \frac{1}{\sqrt 2}(z_k,1,i,-z_k).
	\label{pol}
\end{align}
The definition of $*F$ \eqref{Fdual} and $*F=d\tilde A$ implies
\begin{align}
	\p_\mu \tilde A_\nu - \p_\nu \tilde A_\mu = {\vep_\mn}^\rs \p_\rho A_\sigma. 
	\label{dual}
\end{align}
The vector field $\tilde A_\mu$ is essentially the photon field with a different polarization vector \cite{Strominger:2015bla},
	so let us write the ansatz
\begin{align}
	\tilde A_\mu = \int \td{k} \left(
			\tilde \ep^{\alpha*}_\mu(\V k) a_\alpha(\V k)e^{ik\cdot x}
			+ \tilde \ep^{\alpha}_\mu(\V k) a^{\dagger}_\alpha(\V k)e^{-ik\cdot x}
		\right),
	\label{expansion}
\end{align}
for some polarization vectors $\tilde \ep^\alpha_\mu$.
Then equation \eqref{dual} boils down to
\begin{align}
	k_\mu\tilde \ep^{\alpha*}_\nu(\V k) - k_\nu \tilde \ep^{\alpha*}_\mu(\V k)
		= {\vep_\mn}^\rs k_\rho \ep^{\alpha*}_\sigma.
	\label{eq}
\end{align}
Now, let us make the choice
\begin{align}
	\tilde \ep^\pm_\mu(\V k) = -A_\mn \ep_\pm^\nu(\V k), \qquad A_\mn\equiv \ep_\mnrs \frac{n^\rho k^\sigma}{n\cdot k},
	\label{general}
\end{align}
where $n^\mu\neq k^\mu$ is an arbitrary non-zero 4-vector of our choice.
It is straightforward to check that this constitutes an infinite number of solutions for \eqref{eq}.
An illuminating choice is $n^\mu=(1,0,0,-1)$, for which we obtain
\begin{align}
	A_\mn &= \begin{bmatrix}
		0&-\frac{i}{2}(\zb_k-z_k)&\frac{1}{2}(\zb_k+z_k)&0\\
		\frac{i}{2}(\zb_k-z_k)&0&-1&\frac{i}{2}(\zb_k-z_k)\\
		-\frac{1}{2}(\zb_k+z_k)&1&0&-\frac{1}{2}(\zb_k+z_k)\\
		0&-\frac{i}{2}(\zb_k-z_k)&\frac{1}{2}(\zb_k+z_k)&0
	\end{bmatrix},
\end{align}
and accordingly, the polarization vectors become
\begin{align}
	\tilde \ep^{+\mu}(\V k) &= \frac{1}{\sqrt 2}(-i\zb_k,-i,1,i\zb_k) = -i\ep^{+\mu}(\V k),
	\\
	\tilde \ep^{-\mu}(\V k) &= \frac{1}{\sqrt 2}(iz_k,i,-1,-iz_k) = i\ep^{-\mu}(\V k).
\end{align}
One can see that this $\tilde \ep^\pm$ can essentially be obtained by a $\frac{\pi}{2}$-rotation of $\ep^\pm$ in the complex plane,
	reflecting the electromagnetic duality $\V E\to \V B$ and $\V B\to -\V E$.

Having obtained a solution for $\tilde A$, we now use the methods of Wilson line dressing construction \cite{Jakob:1990zi,Choi:2018oel,Choi:2019fuq}
	to derive the 't Hooft line dressing $\tilde W(p)$.
We first write the dressing \eqref{full} along a path $\Gamma$ at the asymptotic future,
\begin{align}
	\tilde W
	&=
		\exp\left\{
			ig\int_{\Gamma} d\xi^\mu \tilde A_\mu(\xi)
		\right\}
\end{align}
Now we assert that $\Gamma$ is the straight line geodesic of an asymptotic particle with momentum $p$, for which
	we may parametrize $\xi^\mu = \xi_0^\mu + \frac{p^\mu}{m_\chi}\tau$.
Then, using \eqref{expansion} and \eqref{general},
\begin{align}
	\tilde W(p)
	&=
		\exp\left\{
			ig\int^t d\tau\,\frac{p^\mu}{m_\chi} \tilde A_\mu\left(\xi_0+\frac{p}{m_\chi}\tau\right)
		\right\}
	\\ &=
		\exp\left\{
			- g
			\int\td{k}
			\frac{A^\mn p_\nu}{p\cdot k}
			\left(
				\ep_\mu^\alpha(\V k) a_\alpha^\dagger(\V k)e^{ik\cdot (\xi_0+\frac{p}{m_\chi}t)}
				- \ep_\mu^{\alpha *}(\V k)a_\alpha(\V k)e^{ik\cdot (\xi_0+\frac{p}{m_\chi}t)}
			\right)
		\right\},
\end{align}
where we used the boundary condition $\int^t d\tau e^{i\frac{k\cdot p}{m_\chi}\tau}=\frac{m_\chi}{ik\cdot p}e^{i\frac{k\cdot p}{m_\chi}t}$,
	see \cite{Kulish:1970ut} for a discussion.
For an asymptotic particle $t$ diverges to infinity, so by virtue of the Riemann-Lebesgue lemma non-zero frequency modes do not
	contribute to the integral.
To reflect this, we write
\begin{align}
	\tilde W(p)
	&=
		\exp\left\{
			- g
			\int\td{k}
			\frac{p^\mu}{p\cdot k}\phi(\w)
			\left(
				\tilde \ep_\mu^\alpha a_\alpha^\dagger(\V k)
				- \tilde \ep_\mu^{\alpha *}a_\alpha(\V k)
			\right)
		\right\},
	\label{W}
\end{align}
where $\phi(\w)$ is any smooth infrared function \cite{Kulish:1970ut, Ware:2013zja}
	that satisfies $\phi(0)=1$ and has support only in a small neighborhood of $\w=0$.

One can immediately see that this is the dressing $e^{\tilde R(p)}$ we obtained in \eqref{singleFK} from the
	asymptotic $A\bar\chi\chi$ interaction potential,
\begin{align}
	\tilde W(p) = e^{\tilde R(p)}.
	\label{welp}
\end{align}
Therefore the magnetic Faddeev-Kulish dressing may be written as a 't Hooft line dressing.
This is analogous to the electric counterpart associated with Wilson line dressing \cite{Jensen:1995qv}.
These results further justify the choice of an effective large-time interaction of the form \eqref{Vas},
	irrespective of the validity of the full theory \eqref{Lagrangian1}.

\subsection{Soft LGT charges}

We will now show that the 't Hooft line dressing \eqref{W} carries a definite soft magnetic LGT charge.
The soft part of the magnetic LGT charge on $\scri^+$ has the expression \cite{Strominger:2015bla} (up to different normalization)
\begin{align}
	\tilde Q_\vep^+ = i\int d^2z \left(\p_z \vep(z,\zb) F_\zb^+ - \p_\zb \vep(z,\zb) F_z^+\right),
	\label{charge}
\end{align}
where $\vep(z,\zb)$ is a 2-sphere function parametrizing the LGT, which we assume does not introduce poles or branch cuts.
The soft photon operator $F^+_z$ is defined as
\begin{align}
	F^+_z = \int_{-\infty}^\infty du\, F_{uz}^{(0)}
		= \lim_{r\to \infty}\int_{-\infty}^\infty du\, \p_u A_z(u,r,z,\zb).
\end{align}
The field $A_z$ in the large $r$ limit takes the form (see for example Appendix A of \cite{Gabai:2016kuf})
\begin{align}
	\lim_{r\to \infty} A_z(u,r,z,\zb) =
		-\frac{i}{8\pi^2}\sqrt{\gzz}\int_0^\infty d\w\left(a_+(\w \V x_z) e^{-i\w u}-a_-^\dagger(\w \V x_z)e^{i\w u}\right)
	\label{Az}
\end{align}
where $\V x_z$ is the unit 3-vector corresponding to the direction $(z,\zb)$, with components
\begin{align}
	\V x_z = \left(\frac{\zb+z}{1+z\zb},\frac{i(\zb-z)}{1+z\zb},\frac{1-z\zb}{1+z\zb}\right).
\end{align}
Substituting \eqref{Az} into \eqref{charge} and using $\int_{-\infty}^\infty du\,\p_u e^{\pm i\w u}= \pm 2\pi i\w \delta(\w)$,
we obtain
\begin{align}
	\tilde Q_\vep^+ =
		\frac{i}{4\pi}\int d\w d^2z\sqrt{\gzz}\,\w\delta(\w)
		\bigg[&
			\p_\zb \vep(z,\zb) \left (a_-^\dagger(\w\V x_z)+a_+(\w\V x_z)\right )
			-\hc
		\bigg],
\end{align}
where the presence of the delta function shows that only zero-frequency photon operators contribute (hence the soft charge).

Using this expression and the canonical commutation relation \eqref{comm1},
	one can directly compute the commutator of the charge and the operator $\tilde R(p)$ of \eqref{welp}, i.e.,
\begin{align}
	\left[\tilde Q_\vep^+,\tilde R(p)\right] &=
		-\frac{ig}{4\pi}\int d^2z \sqrt{\gzz}\frac{A^\mn p_\nu}{p\cdot \hat k}
		\left(
			\ep^+_\mu(\V x_z)\p_\zb \vep
			-\ep^-_\mu(\V x_z)\p_z\vep
		\right),
\end{align}
where $k^\mu = \w\hat k^\mu$, and we have used the convention \cite{Gabai:2016kuf, Choi:2017bna}
\begin{align}
	\int_0^\infty d\w\,\delta(\w)f(\w) = \frac{1}{2}f(0).
	\label{convention}
\end{align}
By partial integration, this can be put in the form
\begin{align}
	\left[\tilde Q_\vep^+,\tilde R(p)\right] &=
		\frac{ig}{4\pi}\int d^2z\,\vep(z,\zb)
		\left[
			\p_\zb
			\left(
				\sqrt\gzz\,\frac{\ep^+\cdot A\cdot p}{p\cdot \hat k}
			\right)
			- \p_z
			\left(
				\sqrt\gzz\,\frac{\ep^-\cdot A\cdot p}{p\cdot \hat k}
			\right)
		\right]
	\\ &=
		-\frac{g}{2\pi}\int d^2z\,\vep(z,\zb)
		\ \im\left[
			\p_\zb
			\left(
				\sqrt\gzz\,\frac{\ep^+\cdot A\cdot p}{p\cdot \hat k}
			\right)
		\right],
		\label{Q2R2}
\end{align}
where we used the notation $\ep^+\cdot A\cdot p=\ep^+_\mu A^\mn p_\nu$.
Using the identity,
\begin{align}
	\p_\zb\left(
		\sqrt{\gzz}\,\frac{\ep^+\cdot A\cdot p}{p\cdot \hat k}
	\right)
	&=
		-\frac{i}{2}\gzz\left(
			\frac{n^2}{(n\cdot \hat k)^2}
			-\frac{p^2}{(p\cdot \hat k)^2}
		\right),
	\label{id1}
\end{align}
which shows that the expression of \eqref{Q2R2} in square brackets is purely imaginary for all real vectors $n^\mu$, we obtain the commutator
\begin{align}
	\left[\tilde Q_\vep^+,\tilde R(p)\right]
	&=
		\frac{g}{4\pi}\int d^2z\,\gzz\,\vep(z,\zb)
		\left(
			\frac{m_\chi^2}{(p\cdot \hat k)^2}
			+ \frac{n^2}{(n\cdot \hat k)^2}
		\right),
\end{align}
where we have used $p^2+m_\chi^2=0$.
It follows that the 't Hooft line dressing $\tilde W(p)=\exp \tilde R(p)$ carries a definite soft magnetic LGT charge
	parametrized by the momentum $p$,
\begin{align}
	\left[\tilde Q_\vep^+,\tilde W(p)\right] &=
		\frac{g}{4\pi}\int d^2z\,\gzz\,\vep(z,\zb)
		\left(
			\frac{m_\chi^2}{(p\cdot \hat k)^2}
			+ \frac{n^2}{(n\cdot \hat k)^2}
		\right)
		\tilde W(p).
	\label{QW}
\end{align}
It is worth noting that while the charge eigenvalue in \eqref{QW} has an $n$-dependent term,
	this term does not interfere with magnetic charge conservation since it takes the form $g\times\text{(const)}$,
	and $\sum g_\tin = \sum g_\tout$ for a scattering process.
This term acts as a constant shift in the soft magnetic LGT charge of the state and is unmeasurable.

On the other hand, the 't Hooft line dressing does not carry soft electric LGT charge.
To see this, we note that the soft electric charge takes the form \cite{Strominger:2015bla},
\begin{align}
	Q_\vep^+ &=
		-\int d^2z
		\left(
			\p_z \vep(z,\zb)F_\zb^+
			+ \p_\zb \vep(z,\zb)F_z^+
		\right)
	\\ &=
		\frac{1}{4\pi}\int d\w d^2z\sqrt\gzz\, \w\delta(\w)
		\bigg[
			\p_\zb \vep(z,\zb)\left(a_-^\dagger(\w\V x_z) + a_+(\w\V x_z)\right)
			+ \hc
		\bigg].
\end{align}
A computation similar to the magnetic case shows that the dressing's electric LGT charge
	involves the real part of the expression \eqref{id1},
\begin{align}
	\left[Q_\vep^+,\tilde W(p)\right] 
	&=
		\frac{g}{2\pi}\int d^2z \,\vep(z,\zb)
		\ \re
		\left[
			\p_\zb\left(\sqrt{\gzz}\,\frac{\ep^+\cdot A\cdot p}{p\cdot \hat k}\right)
		\right] \tilde W(p)
	= 0
\end{align}
which vanishes identically.

This is to be contrasted with the electric dressing of QED (Wilson line dressing) $W(p)$, which takes the form \cite{Kulish:1970ut}
\begin{align}
	W(p)
	&=
		\exp\left\{
			- e
			\int\td{k}
			\frac{p^\mu}{p\cdot k}\phi(\w)
			\left(
				\ep_\mu^\alpha a_\alpha^\dagger(\V k)
				- \ep_\mu^{\alpha *}a_\alpha(\V k)
			\right)
		\right\}.
\end{align}
One can show the identity,
\begin{align}
	\p_\zb\left(
		\sqrt\gzz\,\frac{p\cdot \ep^+}{p\cdot \hat k}
	\right)
	&=
		-\frac{1}{2}\gzz\frac{p^2}{(p\cdot \hat k)^2},
	\label{id2}
\end{align}
to obtain the following results,
\begin{align}
	\left[Q_\vep^+,W(p)\right] &=
		\frac{e}{2\pi}\int d^2z\,\vep(z,\zb)
		\ \re
		\left[
			\p_\zb\left(
				\sqrt{\gzz}\,\frac{p\cdot \ep^+}{p\cdot \hat k}
			\right)
		\right]W(p)
	\\ &=
		\frac{e}{4\pi}\int d^2z\,\gzz\,\vep(z,\zb)
		\frac{m_\psi^2}{(p\cdot \hat k)^2}W(p),
	\\
	\left[\tilde Q_\vep^+,W(p)\right] &=
		-\frac{e}{2\pi}\int d^2z\,\vep(z,\zb)
		\ \im\left[
			\p_\zb\left(
				\sqrt{\gzz}\, \frac{p\cdot \ep^+}{p\cdot \hat k}
			\right)
		\right]W(p)
	\\&= 0.
\end{align}
Here we have used $p^2+m_\psi^2=0$, with $m_\psi$ the mass of the electrically charged field $\psi$.
Therefore, the Faddeev-Kulish dressing of QED carries only carries a non-zero electric LGT charge.
The duality between the electric and the magnetic charges and their dressings is now manifest.

Although we have constructed the dressings in a theory that has magnetically charged particles,
	we know from \cite{Freidel:2018fsk} that one can retain the electromagnetic duality on the boundary
	even without bulk degrees of freedom carrying magnetic charge.
Accordingly, we can consider the operators that can be obtained by replacing the momentum $p^\mu$ in the dressing \eqref{W}
	with a constant vector $C^\mu$:
\begin{align}
	\tilde W(C)
	&=
		\exp\left\{
			- g
			\int\td{k}
			\frac{C^\mu}{C\cdot k}\phi(\w)
			\left(
				\tilde \ep_\mu^\alpha a_\alpha^\dagger(\V k)
				- \tilde \ep_\mu^{\alpha *}a_\alpha(\V k)
			\right)
		\right\}.\label{C}
\end{align}
From our construction, we can see that these are 't Hooft line operators along a straight line path at the future null infinity $\scri^+$,
	whose direction is given by $C^\mu$.
The vector $C^\mu$ can be understood as a parameter that encodes the way soft magnetic charge is distributed over the sphere.
The 't Hooft line operators are charged under magnetic LGT and neutral under electric LGT, with a constant charge:
\begin{align}
	\left[\tilde Q_\vep^+,\tilde W(C)\right]
	&=
		\frac{g}{4\pi}\int d^2z\,\gzz\,\vep(z,\zb)
		\left(
			\frac{n^2}{(n\cdot \hat k)^2}
			- \frac{C^2}{(C\cdot \hat k)^2}
		\right)
		\tilde W(C),
	\\
	\left[Q_\vep^+,\tilde W(C)\right] &= 0.
\end{align}
Such operators can be used to translate a vacuum to another vacuum carrying a different soft magnetic LGT charge.

\subsection{Remarks on holomorphic potential and soft theorem}

We have seen that there is an infinite number of choices for the dual field $\tilde A$, parametrized by the 4-vector $n^\mu$.
The relation between $\tilde A_\mu$ and $A_\mu$ can be obtained from the relation between polarization vectors \eqref{general},
\begin{align}
	\tilde A_\mu = -\ep_\mnrs (n\cdot \p)^{-1}n^\rho \p^\sigma A^\nu.
\end{align}
While this involves unpleasant differential operators, we can expect a more desirable relation between the holomorphic/antiholomorphic
	potentials at $r\to \infty$,
	since in that limit the angular position $(z,\zb)$ will be identified with the angular direction of outgoing momentum $(z_k,\zb_k)$.
By a derivation analogous to \eqref{Az}, we can obtain
\begin{align}
	\tilde A_\mu(u,z,\zb) &\equiv \lim_{r\to \infty}\tilde A_\mu(u,r,z,\zb)
	\\ &=
		-\frac{i}{8\pi^2 r}\int_0^\infty d\w
		\left(
			\tilde \ep_\mu^{\alpha*}(\V x_z)a_\alpha(\w \V x_z) e^{-i\w u}
			-\tilde \ep_\mu^{\alpha}(\V x_z)a_\alpha^\dagger(\w \V x_z) e^{i\w u}
		\right).
\end{align}
It is straightforward to show that while the dual polarization vectors $\tilde \ep^\pm_\mu = -A_\mn \ep^{\pm\nu}$
	are complicated functions of $n^\mu=(n^0,n^1,n^2,n^3)$ with components
\begin{align}
	\tilde \ep^{+0} &=
		\frac{
			i \left(n^0 \left(\zb^2-1\right)+i n^2 \left(\zb^2+1\right)+2 n^3 \zb\right)
		}{
			\sqrt{2} (n^0 z \zb+n^0-n^1z-n^1 \zb+i n^2 z-i n^2 \zb+n^3 (z \zb-1))
		}
	\\
	\tilde \ep^{+1} &=	
		\frac{
			i \left(n^0 \left(\zb^2-1\right)+2 i n^2 \zb+n^3\zb^2+n^3\right)
		}{
			\sqrt{2} (n^0 z \zb+n^0-n^1 z-n^1 \zb+i n^2 z-i n^2 \zb+n^3 (z \zb-1))
		}
	\\
	\tilde \ep^{+2} &=	
		\frac{
			-n^0\left(\zb^2+1\right)+2 n^1 \zb-n^3 \left(\zb^2-1\right)
		}{
			\sqrt{2} (n^0 z\zb+n^0-n^1 z-n^1 \zb+i n^2 z-i n^2 \zb+n^3 (z \zb-1))
		}
	\\
	\tilde \ep^{+3} &=	
		\frac{
			2 i n^0 \zb-i n^1 \left(\zb^2+1\right)+n^2 \left(\zb^2-1\right)
		}{
			\sqrt{2} (n^0 z\zb+n^0-n^1 z-n^1 \zb+i n^2 z-i n^2 \zb+n^3 (z \zb-1))
		},
\end{align}
and $\tilde \ep^{-\mu} = \left(\tilde \ep^{+\mu}\right)^*$,
	their $z/\zb$ components $\tilde \ep^+_z=\tilde\ep^-_\zb=0$, $\tilde\ep^-_z=-\tilde\ep^+_\zb=ir\sqrt{\gzz}$ do not depend on $n^\mu$.
They satisfy $\tilde \ep_{z/\zb}^+ = -i\ep_{z/\zb}^+$, $\tilde \ep_{z/\zb}^-=i\ep_{z/\zb}^-$, which implies the relation
\begin{align}
	\tilde A_z(u,z,\zb) &= i A_z(u,z,\zb),\qquad
	\\
	\tilde A_\zb(u,z,\zb) &= -i A_\zb(u,z,\zb),
\end{align}
in agreement with the identification made in \cite{Strominger:2015bla}.
\footnote{
	The factors $\frac{4\pi}{e^2}$ in \cite{Strominger:2015bla} are due to the different normalizations of the gauge fields.
}
Therefore for any $n$, we arrive at the same complexified transformations that act on the holomorphic potential $A_z$,
\begin{align}
	(\delta_\vep + i\tilde\delta_\vep)A_z &= 2\p_z \vep,
	\\
	(\delta_\vep + i\tilde\delta_\vep)\tilde A_z &= 0.
\end{align}
As expected, the unphysical freedom to choose $n^\mu$ does not affect the results.

Also, at the end of section \ref{sec:FKconstruction} we have seen that, in a dyonic generalization of the theory, the dressing takes the form \eqref{dyon}.
This can written using \eqref{general} as
\begin{align}
	\exp\left\{
		-\int \td{p}\td{k}\rho(\V p)\phi(\w)
		\left(
			\frac{p\cdot \left(e \ep^\alpha + g \tilde \ep^\alpha\right)}{p\cdot k}
			a_\alpha^{\dagger}(\V k)
			- \hc
		\right)
	\right\}.
	\label{dyon2}
\end{align}
This expression is closely related to the soft theorems and ward identities, since one can use the latter to reconstruct dressed states
	as LGT charge eigenstates \cite{Kapec:2017tkm,Choi:2017ylo,Choi:2019rlz} (see also \cite{Gabai:2016kuf,Choi:2017bna}).
In particular, the coefficient of the creation operator $a_\alpha^\dagger(\V k)$ is tied to the soft factor.
In \cite{Strominger:2015bla}, it was conjectured that the leading soft factor
\begin{align}
	\sum_{j\in\tout}\frac{p_j\cdot (e_j\ep^\alpha + g_j\tilde \ep^\alpha)}{p_j\cdot k}
	- \sum_{i\in\tout}\frac{p_i\cdot (e_i\ep^\alpha + g_i\tilde \ep^\alpha)}{p_i\cdot k}
\end{align}
is exact for all abelian gauge theories.
We emphasize that this result, like the dressing we have constructed, depends only on the asymptotic form of the three-point
	interaction, \eqref{Vas}.
The full content of the theory \eqref{Lagrangian1} is not needed.

\section{Gravitational 't Hooft line dressing}\label{sec:gravity}

In this section, we will consider perturbative quantum gravity in asymptotically flat spacetimes and investigate how the
	construction of 't Hooft line dressings can be extended to the gravitational context.
Since we do not have a Lagrangian field theory to guide us, we proceed by analogy with electromagnetism.

\subsection{Preliminaries}

Consider the asymptotically flat metric on $\scri^+$ in Bondi coordinates \cite{Strominger:2017zoo}
\begin{align}
	ds^2 &= -du^2 - 2dudr + 2r^2\gzz dz^2 d\zb^2
		\nonumber \\&\quad
		+\frac{2m_B}{r}du^2 + rC_{zz}dz^2 + rC_{\zb\zb}d\zb^2 - 2U_z dudz - 2U_\zb dud\zb+\cdots,
		\label{metric2}
\end{align}
where the first line is the Minkowski metric,
	$m_B(u,z,\zb)$ is the Bondi mass aspect, $C_{zz}(u,z,\zb)$ is the radiative mode and $U_z = \frac{1}{2}D^zC_{zz}$.
Here $D_z$ denotes covariant derivative on the 2-sphere (i.e. with respect to $\gzz$).

Let us define the graviton field $h_\mn(x)$ to be
\begin{align}
	g_\mn(x) = \eta_\mn + \kappa h_\mn(x),\qquad \kappa^2=32\pi G.
\end{align}
This relates the radiative gravitons on $\scri^+$ to the metric \eqref{metric2}.
These gravitons have the mode expansion
\begin{align}
	h_\mn(x) = \int \td{k} \left(
			\ep^{\alpha *}_\mn(\V k) a_\alpha(\V k)e^{ik\cdot x} + \ep^\alpha_\mn(\V k) a_\alpha^\dagger(\V k)e^{-ik\cdot x}
		\right),
\end{align}
where we demand the canonical commutation relation
\begin{align}
	\left[a_\alpha(\V k),a_\beta^\dagger(\V k')\right] = \delta_\ab (2\pi)^3(2\w)\delta^{(2)}(\V k-\V k').
	\label{grav_comm}
\end{align}
We are implicitly summing over the physical polarizations $\alpha=\pm$.
Parametrizing graviton momentum $k^\mu$ as in \eqref{photon_momentum}, the graviton polarization vector $\ep^\pm_\mn(\V k)$
	can be written in terms of the two photon polarization vectors \eqref{pol} as $\ep^\pm_\mn(\V k) = \ep^\pm_\mu(\V k)\ep^\pm_\nu(\V k)$.

\subsection{Construction of dressing}

In general relativity, we do not have the magnetic counterpart to the local source term $h_\mn T^\mn$, so we do not have
	bulk degrees of freedom that carry magnetic BMS charge.
However, from \cite{Freidel:2018fsk}, we know that the electromagnetic duality can still be retained on the boundary in this case.
Following this perspective, our goal in this section is to obtain generic operators charged under dual supertranslations.
We achieve this by analogy with electromagnetism: we first construct gravitational 't Hooft line dressings of particles,
	and then replace the particle momentum with a constant vector as in \eqref{C}.

The gravitational Wilson line along a curve $\Gamma$ takes the form \cite{Choi:2019fuq}
\begin{align}
	\exp\left(-im_0\frac{\kappa}{2}\int_\Gamma dx^\mu h_\mn \frac{dx^\nu}{d\tau}\right),
	\label{grav_wl}
\end{align}
where $m_0$ is a parameter with dimension of mass that is possibly different from the particle mass $m$.
This is basically the Wilson line of QED with the replacement $eA_\mu \to -m(\kappa/2)h_\mn dx^\nu/d\tau$.
For an asymptotic particle the trajectory is a straight line, so $mdx^\nu/d\tau=p^\nu$ is a constant vector.
This implies that the gravitational 't Hooft line dressing can be obtained in the same way as in QED, by the replacement
	$\ep_\mn \to \tilde \ep_\mn$, where
\begin{align}
	\tilde \ep_\mn^\pm(\V k) = \tilde \ep^\pm_\mu(\V k) \ep^\pm_\nu(\V k).
	\label{replacement}
\end{align}
Here $\tilde \ep^\pm(\V k)$ is given by \eqref{general}.
Analogous to the case of QED, by taking $\Gamma$ to be the trajectory of an asymptotic particle with momentum $p$,
	the expression \eqref{grav_wl} becomes the gravitational Wilson line dressing $W_g(p)$,
\begin{align}
	W_g(p) = \exp
		\left\{
			\frac{\kappa}{2}\int\td{k}\frac{p^\mu p^\nu}{p\cdot k}\phi(\w)\left(
				\ep_\mn^{\alpha}(\V k) a_\alpha^\dagger(\V k)
				- \ep_\mn^{\alpha*}(\V k)a_\alpha(\V k)
			\right)
		\right\},
\end{align}
where $\phi(\w)$ is the infrared function.
We obtain the gravitational 't Hooft line $\tilde W_g(p)=\exp\tilde R_g(p)$ dressing by the replacement $\ep_\mn \to \tilde \ep_\mn$, which yields
\begin{align}
	\widetilde W_g(p) &=
		\exp\left\{
			-\frac{\kappa m_0}{2 m}
			\int\td{k}\phi(\w)
			\left(
				\frac{(p\cdot A\cdot \ep^\alpha) (p\cdot \ep^\alpha)}{p\cdot k}a_\alpha^\dagger(\V k)
				- \hc
			\right)
		\right\},
	\label{grav_dressing}
\end{align}
where we denote $p\cdot A\cdot \ep^\alpha = p_\mu A^\mn \ep_\nu^\alpha$.

One can think of the two dressings
\begin{align}
	W_g(p) \qquad\text{and}\qquad \tilde W_g(p)
\end{align}
to be realizing the analog of electromagnetic duality of Freidel and Pranzetti \cite{Freidel:2018fsk} \textit{on the boundary}.
Unlike electromagnetism, where we have the bulk duality between $F_\mn$ and $\tilde F_\mn$,
	here in gravity this duality is not realized in the bulk.
Therefore, we must make a departure from the notion of ``dressing'' magnetically charged particles.
In this sense, we define $p$ not as some particle momentum but rather as a general 4-vector, which we will later see parametrizes how
	the dual supertranslation charge of $\tilde W_g$ is distributed over the sphere.

Unlike the case of photons and Lorentz gauge, in gravity we need more work to show that the dressed states exist in de Donder gauge.
The physical states $\Psi$ of the theory are the ones that satisfy the Gupta-Bleuler condition implementing the de Donder gauge,
\begin{align}
	\left(k^\mu a_\mn(\V k) - \frac{1}{2}k_\nu {a^\mu}_\mu(\V k)\right)\ket{\Psi}.
\end{align}
It is straightforward to see that the commutator,
\begin{align}
	&\left[
		\frac{A^{\alpha\rho}p_\rho p^\beta}{p\cdot k}\left(a_\ab^\dagger(\V k)-a_\ab(\V k)\right),
		k'^\mu a_\mn(\V k') - \frac{1}{2}k'_\nu {a^\mu}_\mu(\V k')
	\right]
	=
		\frac{1}{2}p^\mu A_\mn(2\pi)^3(2\w)\delta^{(3)}(\V k-\V k'),
\end{align}
does not vanish, which implies that the dressing in its current form is incompatible with de Donder gauge.
Therefore, just as for the Wilson line dressing \cite{Ware:2013zja,Choi:2017bna}, we must introduce a symmetric tensor $\tilde c_\mn(p,k)$ to fix this.
Let us re-define the dressing to be
\begin{align}
	\tilde W_g(p) = \exp\left\{\frac{\kappa m_0}{2m}\int\td{k}\phi(\w)
		\left(\frac{A^{\mu\rho}p_\rho p^\nu}{p\cdot k}+\frac{1}{\w}\tilde c^\mn(p,k)\right)\left(a_\mn^\dagger(\V k)-a_\mn(\V k)\right)\right\}.
\end{align}
It was shown in \cite{Choi:2017bna} that, in order for this correction to just be a unitary transformation of \eqref{grav_dressing}
	and not introduce additional singularities, we require the condition
\begin{align}
	\tilde c_\mn(p',k)I^\mnrs \tilde c_\rs(p,k) = O(k) \qquad \text{for all $p,p'$,}
	\label{cond1}
\end{align}
where $I_\mnrs = \eta_{\mu\rho}\eta_{\nu\sigma}+\eta_{\nu\rho}\eta_{\mu\sigma}-\eta_\mn\eta_\rs$.
For the states to be well-defined under the Gupta-Bleuler condition, we require
\begin{align}
	0 &=
		\left[
			\left(\frac{A^{\alpha\rho}p_\rho p^\beta}{p\cdot k}+\frac{\tilde c^\ab}{\w}\right)
			\left(a_\ab^\dagger(\V k)-a_\ab(\V k)\right),
			k'^\mu a_\mn(\V k') - \frac{1}{2}k'_\nu {a^\mu}_\mu(\V k')
		\right]
	\\ &=
		\left(
			p^\mu A_\mn
			-\frac{2k^\mu\tilde c_\mn}{\w}
		\right)\frac{1}{2}(2\pi)^3(2\w)\delta^{(3)}(\V k-\V k'),
\end{align}
which translates to the condition
\begin{align}
	k^\mu \tilde c_\mn = \frac{\w}{2} p^\mu A_\mn.
	\label{cond2}
\end{align}
The two conditions \eqref{cond1} and \eqref{cond2} are exactly the ones that we encounter for the Wilson line dressing \cite{Choi:2017bna}
	with the replacement
\begin{align}
	p_\nu\ \to\  \tilde p_\nu\equiv -\frac{1}{2}p^\mu A_\mn.
\end{align}
This implies that the solution is exactly that of \cite{Choi:2017bna} with the same replacement:
\begin{align}
	\tilde c_\mn(p,k) &= \frac{\w}{q\cdot k}\left(\frac{\tilde p\cdot k}{q\cdot k}q_\mu q_\nu - q_\mu \tilde p_\nu - q_\nu \tilde p_\mu\right)
		\\ &= \frac{\w}{2(q\cdot k)}\left(p^\alpha A_{\alpha \nu} q_\mu + p^\alpha A_{\alpha\mu}q_\nu\right),
\end{align}
where $q$ is any null vector; a suitable choice is $q^\mu=(1,-\V k)$, which preserves rotational invariance.
One can readily check that the two consistency conditions \eqref{cond1} and \eqref{cond2} are satisfied.
Therefore, by introducing $\tilde c_\mn$ we have a well-defined dressing that preserves the de Donder gauge condition.

Recently it has been shown by Hirai and Sugishita \cite{Hirai:2019gio} that a careful BRST analysis can remove the necessity of
	objects such as $\tilde c_\mn$, at least in QED.
We will henceforth assume that this can be done here as well, since from \cite{Choi:2017bna} we understand that all $\tilde c_\mn$ does
	is shift the soft charges by some function of $p$; the important part was the existence of a solution $\tilde c_\mn$
	to the conditions \eqref{cond1} and \eqref{cond2}, which we have already shown.
We leave the proof along the lines of \cite{Hirai:2019gio} for future work.

\subsection{Dual supertranslation charge}

We will now see that the gravitational 't Hooft lines are charged under dual supertranslations.
A dual supertranslation is parametrized by a 2-sphere function $f(z,\zb)$,
	which we assume does not introduce poles or branch cuts.
Its charge has the expression \cite{Kol:2019nkc}
\begin{align}
	M_f = \frac{2i}{\kappa^2}\int_{\scri^+_-} d^2z \guzz f(z,\zb)\left(D_\zb^2 C_{zz}-D_z^2 C_{\zb\zb}\right).
	\label{M_f}
\end{align}
In general relativity, there is no source term like $h_\mn T^\mn$ for the magnetic case.
This implies that we do not have a hard charge.
Thus under a partial integration, we have no contribution from the future boundary $\scri^+_+$
	and one can cast \eqref{M_f} into a total derivative over $\scri^+$,
\begin{align}
	M_f = -\frac{2i}{\kappa^2}\int_{\scri^+} dud^2z\guzz \left(D_\zb^2 f \p_u C_{zz} - D_z^2 f \p_u C_{\zb\zb}\right),
\end{align}
where we also integrated by parts twice on the sphere.
Using the mode expansion of $h_{zz}$, the radiative mode $C_{zz}$ can be written as \cite{Choi:2017bna}
\begin{align}
	C_{zz}(u,z,\zb) &= \kappa \lim_{r\to \infty} \frac{1}{r}h_{zz}(u,r,z,\zb)
		\\ &=
			\frac{i\kappa}{8\pi^2}\gzz\int_0^\infty d\w \left(
				a^\dagger_-(\w \V x_z)e^{i\w u} - a_+(\w \V x_z)e^{-i\w u}
			\right).
\end{align}
Using the identity $\int_{-\infty}^\infty du\p_u e^{\pm i\w u} = \pm 2\pi i \w\delta(\w)$, we may write
\begin{align}
	M_f &=
		\frac{i}{2\pi \kappa}\int d\w d^2z\,\w\delta(\w)
		\left[
			D_\zb^2 f \left (a_-^\dagger(\w\V x_z) + a_+(\w\V x_z)\right )
			- \hc
		\right].
		\label{M}
\end{align}
Writing the gravitational 't Hooft line dressing as $\tilde W_g(p) = \exp \tilde R_g(p)$
	and using the canonical commutation relation \eqref{grav_comm},
	one can obtain
\begin{align}
	\left[M_f,\widetilde R_g(p)\right] &=
		\frac{im_0}{4\pi m}\int d^2z
		\Bigg(
			D_\zb^2 f\frac{(\ep^+\cdot A\cdot p)(p\cdot \ep^+)}{p\cdot \hat k}
			- D_z^2 f\frac{(\ep^-\cdot A\cdot p)(p\cdot \ep^{-})}{p\cdot \hat k}
		\Bigg).
\end{align}
Noting that $\Gamma^z_{zz}=\frac{-2\zb}{1+z\zb}=\guzz\p_z \gzz$ and integrating by parts, we can write this in the form
\begin{align}
	\left[M_f,\widetilde R_g(p)\right] &=
		\frac{i m_0}{4\pi m}\int d^2z\,f(z,\zb)
		\left[
			\p_\zb \p^z
			\left(
				\gzz\,\frac{(\ep^+\cdot A\cdot p)(p\cdot \ep^+)}{p\cdot \hat k}
			\right)
			- \cc
		\right]
	\\ &=
		-\frac{m_0}{2\pi m}\int d^2z\,f(z,\zb)
		\ \im
		\left[
			\p_\zb \p^z
			\left(
				\gzz\,\frac{(\ep^+\cdot A\cdot p)(p\cdot \ep^+)}{p\cdot \hat k}
			\right)
		\right],
		\label{MfRg}
\end{align}
where we denote $\p^z = \guzz \p_\zb$.
One can show that
\begin{align}
	\p_\zb\p^z\left(\gzz\,\frac{(\ep^+\cdot A\cdot p)(p\cdot \ep^+)}{p\cdot \hat k}\right)
	&=
		-\frac{i}{2}\gzz\left(
			\frac{m^4}{(p\cdot \hat k)^3}
			- \frac{n^2}{(n\cdot \hat k)^3}B(p,n,z)
		\right),
\end{align}
where
\begin{align}	
	B(p,n,z) &=
		(n^1+in^2)(p^1-ip^2)
		-(n^0-n^3)(p^0+p^3)
		\nonumber\\&\quad
		-(n^0+n^3)(p^1-ip^2)z
		+(n^1-in^2)(p^0+p^3)z.
\end{align}
Unlike the case of electromagnetism,
	one can see that an obscure choice of $n$ can make the expression in square brackets of \eqref{MfRg} contain both imaginary and real components.
A sufficient condition to prevent this is to choose $n$ to be null;
	an example is $n^\mu = (1,0,0,-1)$ which, as we saw in section \ref{sec:dressing_as_tHooft_line},
	corresponds to a $\frac{\pi}{2}$-rotation of $\ep^\pm$ in the complex plane.
Then, we have the commutator
\begin{align}
	\left[M_f,\tilde W_g(p)\right] &= \left[M_f,e^{\tilde R_g(p)}\right] \\&=
		\left(
			\frac{m_0}{4\pi m}\int d^2z\,\gzz\,f(z,\zb)
			\frac{p^4}{(p\cdot \hat k)^3}
		\right)\tilde W_g(p),
\end{align}
which implies that the gravitational 't Hooft line dressing \eqref{grav_dressing} carries dual supertranslation charge.

On the other hand, it does not carry supertranslation charge.
To see this, we first recall that the soft supertranslation charge $T_f$ has the form \cite{He:2014bga,Choi:2017bna}
\begin{align}
	T_f = \frac{1}{2\pi\kappa}\int d\w d^2z\,\w\delta(\w)\left[
			D_\zb^2 f\left(
				a_-^\dagger(\w\V x_z)
				+a_+(\w\V x_z)
			\right)
			+\hc
		\right].
\end{align}
Then, by a similar analysis, one obtains,
\begin{align}
	\left[M_f,\widetilde W_g(p)\right] &=
		\frac{1}{2\pi}\int d^2z\,f(z,\zb)
		\ \re
		\left[
			\p_\zb \p^z
			\left(
				\gzz\,\frac{(\ep^+\cdot A\cdot p)(p\cdot \ep^+)}{p\cdot \hat k}
			\right)
		\right]\tilde W_g(p)=0,
\end{align}
which proves the statement.

It is instructive to contrast this with the gravitational Faddeev-Kulish dressing (gravitational Wilson line dressing), which takes the form
	\cite{Ware:2013zja}
\begin{align}
	W_g(p) &=
		\exp\left\{
			\frac{\kappa}{2}
			\int\td{k}\phi(\w)
			\frac{p^\mu p^\nu}{p\cdot k}
			\left(
				\ep_\mn^\alpha(\V k)a_\alpha^\dagger(\V k)
				- \hc
			\right)
		\right\}.
\end{align}
The following identity will play a role:
\begin{align}
	\p_\zb \p^z \left(
		\gzz\,\frac{(p\cdot \ep^+)^2}{p\cdot \hat k}
	\right) &=
		\frac{1}{2}\gzz\,\frac{m^4}{(p\cdot \hat k)^3}.
\end{align}
Use of this identity immediately gives:
\begin{align}
	\left[T_f, W_g(p)\right] &=
		\frac{1}{2\pi}\int d^2z\,f(z,\zb)
		\ \re
		\left[
			\p_\zb \p^z
			\left(
				\gzz\,\frac{(p\cdot \ep^+)^2}{p\cdot \hat k}
			\right)
		\right]W_g(p)
	\\ &=
		\left(\frac{1}{4\pi}\int d^2z\,\gzz\,f(z,\zb)
		\frac{m^4}{(p\cdot \hat k)^3}\right)
		W_g(p),
	\\
	\left[M_f, W_g(p)\right] &=
		-\frac{1}{2\pi}\int d^2z\,f(z,\zb)
		\ \im
		\left[
			\p_\zb \p^z
			\left(
				\gzz\,\frac{(p\cdot \ep^+)^2}{p\cdot \hat k}
			\right)
		\right]W_g(p)
	\\ &= 0,
\end{align}
implying that the gravitational Wilson line dressing only carries a definite supertranslation charge.

Since we do not have a ``magnetic'' counterpart to the source term $h_\mn T^\mn$ in general relativity,
	we have no bulk degrees of freedom carrying dual supertranslation charge.
Therefore, instead of the dressing \eqref{grav_dressing} of ``magnetically'' charged particles, we generalize them to generic operators
	that are charged under dual supertranslation.
These operators can be obtained by replacing $p^\mu$ with a constant vector $C^\mu$ as in \eqref{C}
	and absorb the dimensionless factor $m_0/m$ in $C^\mu$:
\begin{align}
	\widetilde W_g(C) &=
		\exp\left\{
			-\frac{\kappa}{2}
			\int\td{k}\phi(\w)
			\left(
				\frac{(C\cdot A\cdot \ep^\alpha) (C\cdot \ep^\alpha)}{C\cdot k}a_\alpha^\dagger(\V k)
				- \hc
			\right)
		\right\}.
\end{align}
From our construction, we see that these are 't Hooft line operators along a straight line geodesic at $\scri^+$ whose direction is
	given by $C^\mu$.
The vector $C^\mu$ parametrizes how the soft dual supertranslation charge is distributed over the 2-sphere.
Choosing $n$ to be null, we see that
\begin{align}
	\left[M_f,\tilde W_g(C)\right] &=
		\left(
			\frac{1}{4\pi}\int d^2z\,\gzz\,f(z,\zb)
			\frac{C^4}{(C\cdot \hat k)^3}
		\right)\tilde W_g(C),
	\\
	\left[T_f,\tilde W_g(C)\right] &= 0.
\end{align}
These 't Hooft line operators are charged under dual supertranslation and neutral under supertranslation,
	and we can use them to translate a vacuum to another vacuum carrying a different dual supertranslation charge.

\section{Discussion}\label{sec:discussion}

In a quantum field theory of electric and magnetic charges, we have constructed the asymptotic states by following
	the original method of Faddeev and Kulish.
We have shown that the magnetic dressings can be expressed as a 't Hooft line operator, and that they are charged under
	magnetic large gauge transformations.
The 't Hooft line interpretation allowed us to formulate a gravitational 't Hooft line operator, which is charged under dual supertranslations
	while carrying zero supertranslation charge.

In this work, we have assumed the 2-sphere function parameters $\vep(z,\zb)$ and $f(z,\zb)$ to be smooth.
One can, however, consider the more general case where they contain poles.
Then, the soft electric and magnetic charges do not commute, and their algebra possesses a central charge
	\cite{Hosseinzadeh:2018dkh, Freidel:2018fsk}.
To see this, for instance in electromagnetism, one can consider the boundary fields
\begin{align}
	a_{z/\zb}(z,\zb) &\equiv A^+_{z/\zb}(z,\zb) - A^-_{z/\zb}(z,\zb), \\
	\tilde a_{z/\zb}(z,\zb) &\equiv \tilde A^+_{z/\zb}(z,\zb) - \tilde A^-_{z/\zb}(z,\zb),
\end{align}
where $A^\pm_{z/\zb}(z,\zb)\equiv \lim_{u\to\pm\infty}A_{z/\zb}(u,z,\zb)$ and similarly for $\tilde A_{z/\zb}$'s.
They satisfy $[a_z,\tilde a_\wb]=i\delta^{(2)}(z-w)$, which is derived using \eqref{comm1}
	with a correction factor of 2 due to indistinguishability of helicity at zero momentum;
	see \cite{He:2014cra} for a rigorous treatment.
Then, one can write the soft charges as
\begin{align}
	Q^+_{\vep_1} &= -i\int_{S^2} d\vep_1 \wedge \tilde a,
	\\
	\tilde Q^+_{\vep_2} &= i\int_{S^2} d\vep_2 \wedge a.
\end{align}
In the presence of poles, the surface $S^2$ will contain circular boundaries around the poles and will lead a central charge
	\cite{Hosseinzadeh:2018dkh,Freidel:2018fsk}
\begin{align}
	\left[Q^+_{\vep_1},\tilde Q^+_{\vep_2}\right]
	&=
		i\int_{S^2} d\vep_1\wedge d\vep_2
	= -i\sum_p \oint_p \vep_1\, d\vep_2.
\end{align}
Since the soft electric and magnetic charges do not commute, we cannot label the states with both charges simultaneously.
Instead, we can label the states using electric charges with poles in $\vep(z,\zb)$.
The states with magnetic LGT charges are then written as a superposition of electrically charged states \cite{Freidel:2018fsk}.
This is therefore a purely quantum phenomenon.
It will be very interesting to see how this will affect the construction of dressings.
We leave this for future investigation.

Another future direction is the extension of the magnetic dressing construction to the Schwarzschild black hole horizon.
In \cite{Choi:2019fuq}, it was shown that one can study the horizon supertranslation hairs by constructing
	dressing associated with BMS supertranslation, and that they involve only the electric parity graviton modes;
	the magnetic parities cancel out in a non-trivial way.
Our work suggests that there should reside another set of soft hair on the horizon, corresponding to dual supertranslations.
It will be interesting to see whether these hairs are associated with the magnetic parity gravitons.
Also, investigating the general case where $f(z,\zb)$ contains poles and studying the implication of the central charge
	would be another interesting and very non-trivial exercise.

Note added:
After the submission of the first version of our paper, we noticed an interesting paper \cite{Henneaux:2020nxi}
	that discusses electromagnetic duality and soft charges.

\begin{acknowledgments}
We are very grateful to Uri Kol for helpful comments and discussions.
SC is supported by the Samsung Scholarship and the Leinweber Graduate Fellowship.
\end{acknowledgments}

\bibliographystyle{apsrev4-2}
\bibliography{magnetic}

\end{document}